\documentclass[journal,twoside,hidelinks,a4paper]{IEEEtran}

\newcommand{\subparagraph}{}

\usepackage{emptypage,titling,titlesec,widetext,textcomp,relsize,parskip}
\usepackage[utf8]{inputenc}
\usepackage[margin=10pt,font=footnotesize,labelsep=period]{caption}
\usepackage{tabulary,longtable,makecell,multirow,color,colortbl}
\usepackage[colorlinks=false,linkcolor=black,urlcolor=gray,citecolor=black]{hyperref}
\usepackage[usenames,dvipsnames,svgnames,table]{xcolor}
\usepackage{authblk,lipsum,txfonts,float,morefloats,gensymb,graphicx,setspace,fancyhdr,bibunits,booktabs,array,marvosym,xhfill}
\usepackage[normalem]{ulem}
\usepackage[export]{adjustbox}

\usepackage{fontawesome}
\usepackage{placeins}
\usepackage{listings} 
\usepackage{rotating} 
\usepackage{pifont}   
\graphicspath{{Figures/}}
\DeclareGraphicsExtensions{.pdf,.png,.bmp,.jpg,.eps}

\definecolor{mycolor}{RGB}{0,104,152}
\definecolor{lightyellow}{rgb}{1.0, 1.0, 0.88}
\definecolor{dkgreen}{rgb}{0,0.6,0}
\definecolor{lightblue}{rgb}{0.0,0.0,0.9}
\definecolor{gray}{rgb}{0.5,0.5,0.5}
\definecolor{mauve}{rgb}{0.58,0,0.82}
\definecolor{gray}{rgb}{0.4,0.4,0.4}
\definecolor{darkblue}{rgb}{0.0,0.0,0.6}
\definecolor{cyan}{rgb}{0.0,0.6,0.6}
\definecolor{darkred}{rgb}{0.6,0.0,0.0}
\definecolor{bole}{rgb}{0.47, 0.27, 0.23}

\lstdefinelanguage{XML}
{
	morestring=[s][\color{mauve}]{"}{"},
	morestring=[s][\color{black}]{>}{<},
	morecomment=[s]{<?}{?>},
	morecomment=[s][\color{dkgreen}]{<!--}{-->},
	stringstyle=\color{black},
	identifierstyle=\color{lightblue},	
	keywordstyle=\color{red},
	morekeywords={LOC, NOP, NOC, NOA, NOM, NOCo, NOIn, NOI, NOAc, Name, AccessLevel, isInterface, Superclass, InterfaceName, Type, ReturnType, isStatic, NumberOfParameters, CommentText, Used, LeftHandSide, RightHandSide, ExceptionType, Arguments, InvokedBy, HowIsItUsed, AccessedIn}
}

\usepackage[includeheadfoot,paperwidth=8.5in,paperheight=11in,top=1cm,bottom=2cm,left=2cm,right=2cm,footskip=.25in,headheight=40pt,headsep=12pt,heightrounded]{geometry}

\newcommand{\aka}{\textit{aka.}~}
\newcommand{\etal}{\textit{et}~\textit{al.}~}
\newcommand{\ie}{\textit{i.e.,}~}
\newcommand{\eg}{\textit{e.g.,}~}
\newcommand{\resp}{\textit{resp.}~}

\newcommand{\cf}{\textit{cf.}~}

\newcommand{\xfill}[2][1ex]{{\dimen0=#2\advance\dimen0 by #1\leaders\hrule height \dimen0 depth -#1\hfill}}
\newcommand{\xfilll}[2][1ex]{\dimen0=#2\advance\dimen0 by #1\leaders\hrule height \dimen0 depth -#1\hfill}

\renewcommand{\figurename}{Figure}

\titlespacing\section{0pt}{8pt plus 4pt minus 2pt}{2pt plus 2pt minus 2pt}
\titlespacing\subsection{0pt}{8pt plus 4pt minus 2pt}{2pt plus 2pt minus 2pt}
\titlespacing\subsubsection{0pt}{8pt plus 4pt minus 2pt}{2pt plus 2pt minus 2pt}

\newcommand{\subtitle}[1]{\posttitle{\par\end{center}\begin{center}\bfseries\Large #1\end{center}\vskip1.2em}}
\newcommand\myabstract[2][.8]{\renewcommand\maketitlehookd{\mbox{}\medskip\par\centering\begin{minipage}{#1\textwidth}\small#2\end{minipage}}}

\definecolor{myred}{RGB}{189, 3, 34}
\date{}

\title{\vspace{-1cm}{\includegraphics[height=0.5in,keepaspectratio=true]{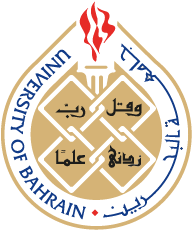}\hfill\parbox{15cm}{\raggedleft\sublargesize{\color{myred}\bfseries{International Journal of Computing and Digital Systems\\}}\sublargesize{2025, VOL. 17, NO.1, 1-20\\\vspace{3mm} \url{http://dx.doi.org/10.12785/ijcds/1571046420}}}\\\vspace{1cm}{\bfseries{\fontsize{18}{21.6}ScaMaha: A Tool for Parsing, Analyzing, and Visualizing Object-Oriented Software Systems}}}}

\subtitle{}

\author[1]{\bfseries Ra'Fat Al-Msie'deen}

\affil[1]{\normalfont\textit{Department of Software Engineering, Faculty of IT, Mutah University, Mutah 61710, Karak, Jordan}}
\affil[ ]{\vspace{0pt}}
\affil[ ]{Received 3 July 2024, Revised 5 January 2025, Accepted 9 January 2025}
\affil[ ]{\vspace{0pt}}

\fancypagestyle{firstpage}{%
    \fancyhf{}
    \fancyhead[RO]{{\normalfont{International Journal of Computing and Digital Systems\hspace{4mm}}}\includegraphics[height=0.5in, keepaspectratio=true]{ub_logo}\hspace{3mm}\bfseries{\thepage}}
    \fancyhead[LE]{\bfseries{\thepage}\hspace{3mm}\includegraphics[height=0.5in,keepaspectratio=true]{ub_logo}\hspace{3mm}\normalfont\textit{Ra'Fat Al-Msie'deen}}
    \fancyfoot[R]{\vspace{5mm}\textsl{\color{gray}{\textsf{\url{}}}}}}

\fancypagestyle{plain}{%
    \fancyhf{}
    \fancyhead[LR]{\vspace{2.5cm}}
    \fancyfoot[L]{\vspace{0mm}{\small\textit{E-mail address: rafatalmsiedeen@mutah.edu.jo}}}
    \fancyfoot[R]{\vspace{5mm}\textsl{\color{gray}{\textsf{\url{}}}}}}

\begin{document}
\bstctlcite{IEEEexample:BSTcontrol}
\myabstract[1]{\vspace{-1cm}\hrule\vspace{2mm}\textbf{Abstract:} Reverse engineering tools are required to handle the complexity of software products and the unique requirements of many different tasks, like software analysis and visualization. Thus, reverse engineering tools should adapt to a variety of cases. Static Code Analysis (SCA) is a technique for analyzing and exploring software source code without running it. Manual review of software source code puts additional effort on software developers and is a tedious, error-prone, and costly job. This paper proposes an original approach (called ScaMaha) for Object-Oriented (OO) source code analysis and visualization based on SCA. ScaMaha is a modular, flexible, and extensible reverse engineering tool. ScaMaha revolves around a new meta-model and a new code parser, analyzer, and visualizer. ScaMaha parser extracts software source code based on the Abstract Syntax Tree (AST) and stores this code as a code file. The code file includes all software code identifiers, relations, and structural information. ScaMaha analyzer studies and exploits the code files to generate useful information regarding software source code. The software metrics file gives unique metrics regarding software systems, such as the number of method access relations. Software source code visualization plays an important role in software comprehension. Thus, ScaMaha visualizer exploits code files to visualize different aspects of software source code. The visualizer generates unique graphs about software source code, like the visualization of inheritance relations. ScaMaha tool was applied to several case studies from small to large software systems, such as drawing shapes, mobile photo, health watcher, rhino, and ArgoUML. Results show the scalability, performance, soundness, and accuracy of ScaMaha tool. Evaluation metrics, such as precision and recall, demonstrate the accuracy of ScaMaha in parsing, analyzing, and visualizing software source code, as all code artifacts — including code files, software metrics files, and code visualizations — were correctly extracted.
\newline
\newline
\textbf{Keywords:} Software engineering, Reverse engineering, Software re-engineering, Object-Oriented source code, Static code analysis, Software visualization, Software metrics, ScaMaha tool. \vspace{1mm} \hrule}

\maketitle

\section{INTRODUCTION}
\label{sec:introduction_and_overview}

\begin{figure*}[!htbp]
	\includegraphics[width=\textwidth]{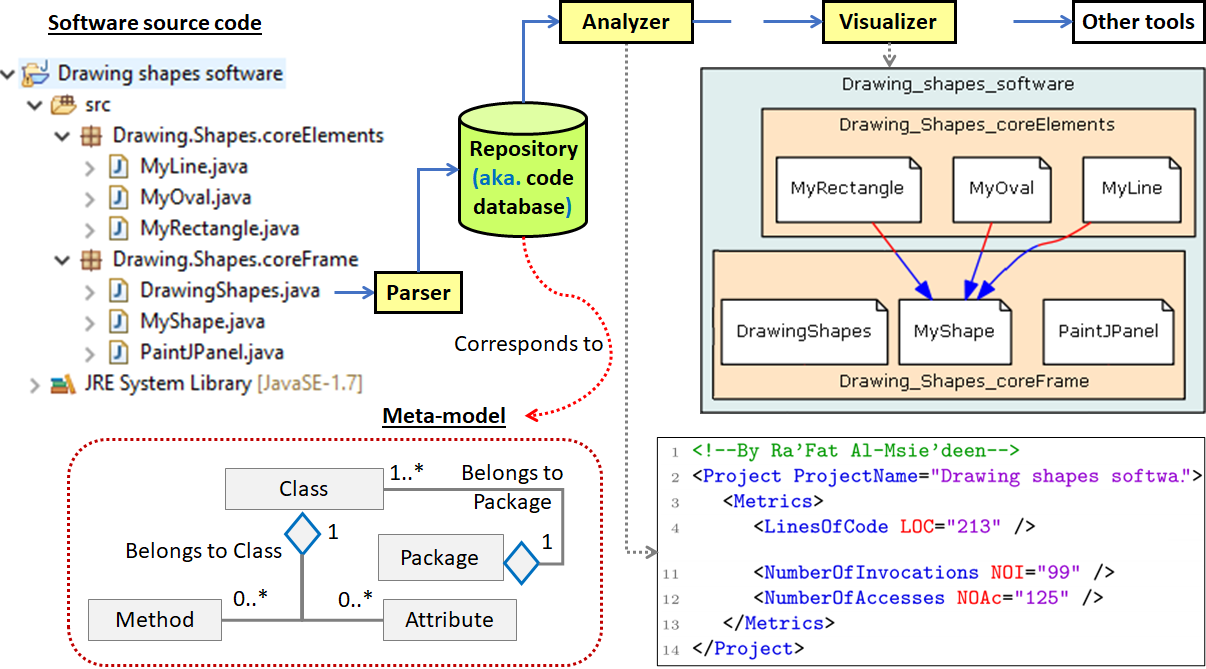}
	\caption{Typical infrastructure for re-engineering tools.}
	\label{Fig:reengineering}
\end{figure*}

Reverse engineering tools are necessary to cope with the complexity of software systems. Also, such tools should cope with the specific requirements of the various reverse engineering tasks, like software comprehension and visualization. Thus, these tools should adapt to a wide range of cases \cite{NicolasAnquetil}. To analyze software code, tools need to represent it. Such a representation should be comprehensive. Software comprehension is still a manual activity. Thus, advanced tools will help developers fully understand complex systems \cite{DBLPrrPanDHKY23}. This paper presents ScaMaha tool, which is a reverse engineering tool for performing software analysis and visualization. The core of ScaMaha tool revolves around the meta-model, code parser, code analyzer, and code visualizer. With ScaMaha, tool developers can analyze and visualize software code. Also, developers can develop new, specific, and dedicated reverse engineering tools based on the infrastructure of ScaMaha tool (\cf Figure \ref{Fig:reengineering}).

Software has become an integral part of modern life, permeating nearly every domain, including smart cities, education, healthcare, robotics, and gaming \cite{NadiaMR}, \cite{JournalAlMsiedeenDB}. In smart cities, software enables the seamless integration of infrastructure, services, and data \cite{RafatSCs}. In education and healthcare, it supports learning systems, patient management, and innovative solutions \cite{RafatSCs2}. In robotics, software drives automation and intelligent task execution, while in gaming, it powers immersive experiences with realistic graphics and AI-driven gameplay. However, the growing complexity of software systems, combined with incomplete documentation and the need to maintain and evolve legacy systems, poses significant challenges for developers. To address these issues, developers rely on advanced tools to manage complexity and to understand, analyze, and visualize legacy code. These tools are essential for ensuring the efficiency and sustainability of software solutions across diverse sectors.

SCA techniques analyze software products by examining their source code without executing them. These techniques improve software quality by identifying potential code faults or vulnerabilities in the early stages of software development \cite{DBLP:PrahoLG1}. SCA is a software verification method aimed at capturing faults in software code early to avoid costly repairs later \cite{Ilyas947354}. Nowadays, SCA is broadly used, and several tools are freely available for different programming languages, such as Java, C++, and others. SCA is the method of exploring software code without running it to find the main code elements (\eg class and method names) and their relations (\eg inheritance and method invocation) \cite{DBLPTrautschEHG23, AlMsieDeen1445}. In contrast, dynamic code analysis involves running the software code and observing how it behaves and performs while it runs \cite{Desilva2023}.

In this work, the subject of study is the source code of OO software systems. More precisely, the author is interested in parsing, analyzing, and visualizing software source code. Software Engineering is the systematic way to build and maintain software systems \cite{DBLP0017996}. Software source code is considered as one of the important resources from which software system is built \cite{JurgenJordanusVinju}. Moreover, reverse engineering aims at analyzing legacy software systems to retrieve their design and other information based on their software source code \cite{DBLPMenguy23}.

The main challenge in evolving and maintaining legacy software products is comprehending the chosen software. Reverse engineering is the process of studying and analyzing software products \cite{AtiJain}. The goal of this process is to identify the product's components and their relationships. Furthermore, this process aims at creating several representations of the software product in another shape or at a different level of abstraction. The main goal of the software re-engineering process is to improve or modify current software so it can be comprehended, administered, and used again as new software \cite{DBLPAssuncaoVL23a, MajthoubManar}.

Software source code analysis presents significant information for the re-engineering and reverse engineering activities of software products. It assists software engineers in software evolution, maintenance, visualization, reuse, and understanding \cite{DBLPMushtaqRS17}. It is anticipated that the volume of software systems in 2025 will exceed 1 trillion Lines of Code (LOC), which shows the value of code analysis in the future \cite{Kitchenham}. The source code of any software can be analyzed through numerous methods. For instance, software code may be analyzed using static or dynamic methods \cite{DBLPAngererGPG19, DBLPAriasAA08}. Also, the software code can be analyzed with a hybrid method that combines static and dynamic methods.

Parsing software source code in order to analyze it is a central activity of many software engineering tasks such as feature location, code summarization, and visualization. Thus, software source code parsing is one of the main software engineering activities. Code parsing is required when a software engineer maintains, visualizes, documents, reuses, migrates, or enhances software systems. When a software developer deals with structural representations of software code, such as in the form of an AST, the process of data pre-processing becomes quite sophisticated and necessitates the use of code parsers (or code analysis tools). An important method of expressing the structure information of software code is AST \cite{DBLP:Zhang}.

Figure \ref{Fig:reengineering} illustrates the typical organization of a re-engineering environment. The left side of Figure \ref{Fig:reengineering} displays the software source code, which can be brought into this environment using suitable code parsers, such as ScaMaha parser. Also, Figure \ref{Fig:reengineering} displays the main repository of this environment, which holds the software code (\aka code database). The repository includes an abstracted model of the software code, which is based on ScaMaha's meta-model. The right side of Figure \ref{Fig:reengineering} displays the tools (\eg ScaMaha analyzer and visualizer) that utilize the repository as their information source to perform specific tasks. The key part that makes all tools work together is the repository's meta-model.

To support software comprehension, visualization, and maintenance, meta-models are often employed during software reverse engineering tasks to describe the components of software and their relationships. Reverse engineering tools frequently define their own meta-models depending on the intended goals and functionalities \cite{HironoriYann}.

Each programming language (\eg Java) has its own rigorous syntax that can be thought of as a collection of predetermined rules revealing all probable programming language constructions. Analyzing the raw software code with these predefined rules allows depicting it in the shape of a parse tree and, after that, as an AST \cite{DBLP:Utkin}. By dealing with this structure, scholars enhance their findings for several software engineering duties, such as code visualization and comprehension \cite{AlkkMsiedeenSHUV201516}. In software engineering domain, tools are typically based on several tools executing particular tasks, such as mining code identifiers, performing code analysis, and visualizing code \cite{DBLP:BellayG97}. Thus, a common code meta-model (\cf Figure \ref{Fig:reengineering}) is required to represent information or facts about the software that is being analyzed \cite{Ducasse}.

This study presents ScaMaha tool, which parses, analyzes, and visualizes OO source code. ScaMaha parser generates an XML file (called a code file) representing software code (\cf Listing \ref{xmlFile}). Software developers can use this code parser in any work that deals with software code, such as feature identification \cite{ALMsiUVICSR, DBLPAlMsiedeenSHUV13, DBLP:Al-MsiedeenSHUVS13} and software evolution \cite{RaFatImpact}. ScaMaha tool extracts all code identifiers and relations.

In addition to code parser, this study presents ScaMaha analyzer. This analyzer accepts as input the software code file that was produced by ScaMaha parser to generate a software metrics file (an XML file). The software metrics file contains quantitative information regarding software source code, like the number of software classes and methods. Software engineers can use or extend the current version of ScaMaha analyzer to examine other aspects of software code. The novelty of this analyzer is that it exploits code information to uniquely identify some features of software code, such as the number of method invocations and attribute accesses. ScaMaha analyzer can be used in several software engineering research areas, such as software maintenance.

\begin{figure}[!htb]
	\center
	\includegraphics[width=\columnwidth]{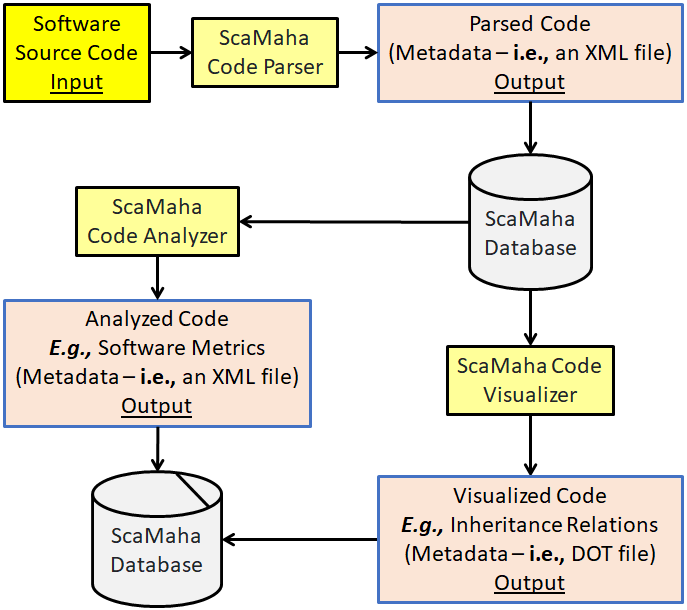}
	\caption{An overview of ScaMaha tool.}
	\label{Fig:useOfSCAMaha}
\end{figure}

In addition to a code parser and analyzer, ScaMaha also presents a code visualizer. This visualizer accepts as input the code file generated via ScaMaha parser and produces a set of code graphs. Each graph addresses a specific aspect of software code. For example, one kind of graph shows the structural information of software code, and another one shows the inheritance relations across software classes. Software visualization is a hot topic in the software engineering domain \cite{DBLPLimaMGPGM23}. Graphs give software developers an indication of software size (or complexity level). In addition, graphs present code information in its simplest form to software developers. The current version of ScaMaha visualizer can be easily extended to include other kinds of graphs covering different aspects of software code. Figure \ref{Fig:useOfSCAMaha} briefly shows the use of ScaMaha tool in this work.

In this study, the only input for ScaMaha is the OO source code. The first step is aimed at parsing software source code statically using the AST. The output produced by ScaMaha code parser (\ie parsed code) can be named as metadata (\ie an XML file called a code file). The parsed code is saved into a code database for further use. The second step of ScaMaha tool is aimed at analyzing the software source code. The output produced via ScaMaha code analyzer can be named as metadata (\ie an XML file called a software metrics file). The analyzed code (\ie software metrics) is kept in the tool database. Finally, the third step of the suggested tool is aimed at visualizing the software source code. Several graphs regarding software code are produced and stored in the tool database (\cf Figure \ref{Fig:useOfSCAMaha}). 

\begin{figure}[!htb]
	\center
	\includegraphics[width=\columnwidth]{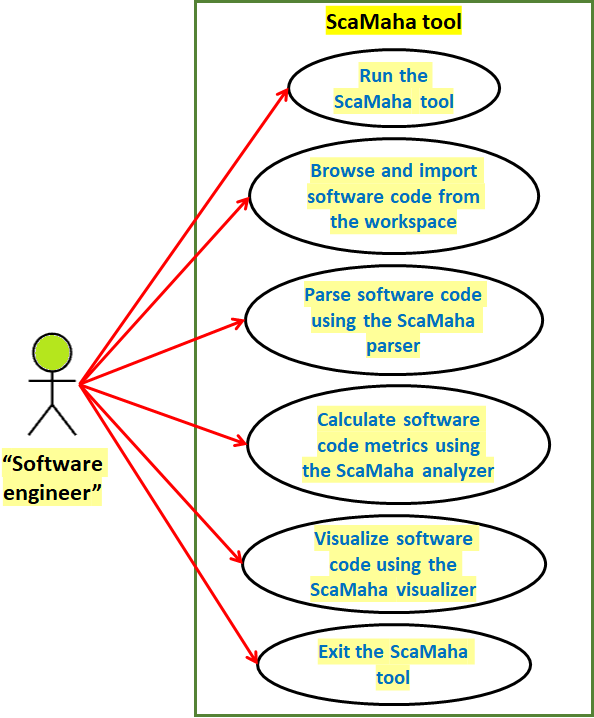}
	\caption{The use-case diagram of ScaMaha tool.}
	\label{Fig:useCase}
\end{figure}

Figure \ref{Fig:useCase} presents the use-case diagram of ScaMaha tool. The use-case diagram shows all possible interactions between external users (\ie software engineers) and ScaMaha. The use-case diagram displays a collection of actions (called use-cases) that are supported by the proposed tool, such as parse and visualize software source code. This diagram defines a collection of use-cases that ScaMaha tool can execute in collaboration with end users to provide them with significant outcomes regarding software code. The source code of ScaMaha, the tutorial, the experimentation results, case studies, and all materials regarding this study are publicly available on ScaMaha web page \cite{ScaMaha, ScaMahaGithub}.

\begin{figure}[!htb]
	\center
	\includegraphics[width=\columnwidth]{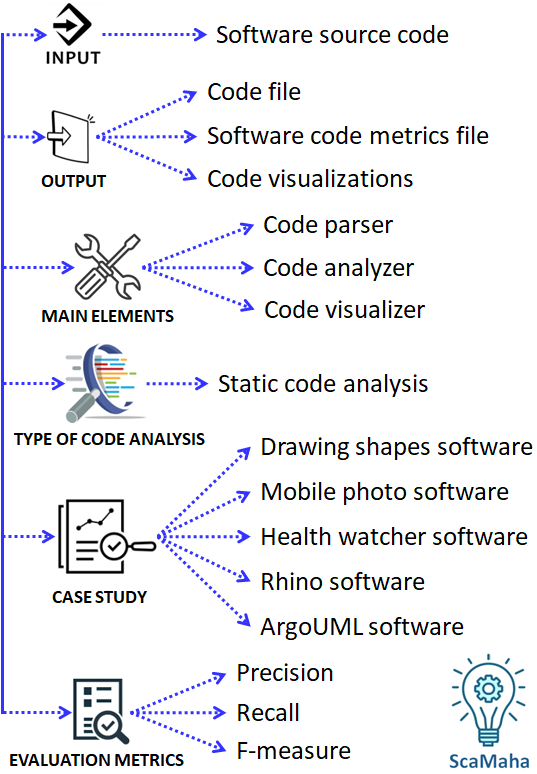}
	\caption{The main elements of ScaMaha approach.}
	\label{Fig:MainEle}
\end{figure}

This paper proposes an automatic approach to analyzing and visualizing OO software systems. ScaMaha tool is the main outcome of this study. ScaMaha is a software engineering tool. What distinguishes ScaMaha tool from others is that it is extensible and can perform many tasks of reverse engineering, such as source code visualization. With ScaMaha, tool developers can develop advanced reverse engineering tools that can exploit the existing components of ScaMaha, like the meta-model and code parser. Figure \ref{Fig:MainEle} illustrates the main elements of ScaMaha approach.

The rest of this paper is arranged as follows: Current studies related to ScaMaha contributions are presented in Section \ref{sec:related_work}. ScaMaha is detailed in Section \ref{sec:Approach}. Experiments are given and discussed in Section \ref{sec:Exper.}. Finally, Section \ref{sec:conclusion} concludes this study and provides proposals for future work.

\setlength{\parindent}{5mm}

\section{Related Work} \label{sec:related_work}

This section offers a literature review associated with ScaMaha contributions. The closest works to ScaMaha are chosen and offered in this section.

Bruneliere \etal \cite{DBLPBruneliereCJM10} suggested a semantic and syntax analysis based parser for the creation of AST and metrics for multi-language software systems. Their meta-modeling tool \cite{MoDisco} is utilized to analyze multi-language applications \cite{DBLPBruneliereCDM14}.

VerveineJ is a parser developed in Java that constructs an MSE file from software source code \cite{VerveineJTechnology}. Based on Eclipse Java Development Tools (JDT), VerveineJ parses software code written in Java to export it in the MSE format that is utilized by the Moose data analysis platform \cite{VerveineJMoose, Moose}. Moreover, VerveineJ is used to extract relationships (or dependencies) from software source code.

Janes \etal \cite{DBLPJanesPSS13} proposed an uncommon open-source solution that prevents producing parsers from scratch or dealing with parser generators. They proposed and described how to employ parsers included in the Eclipse Integrated Development Environment (IDE) \cite{EclipseIDE} to parse software code, such as the JSDT parser \cite{JSDT}.

Parsing and analyzing software source code is a critical part of the reverse engineering process. Nowadays, several source code parsers exist. Software engineers can utilize those parsers to deal with numerous software engineering activities, like software comprehension and visualization. O'Hara and Slavin \cite{DBLPOHaraS19} described in their work a collection of tools for parsing and analyzing many different programming languages, involving legacy languages like Fortran. Table \ref{tab:ScaMahaChara} displays the main characteristics of ScaMaha tool.

\begin{table}[!htbp]
	\center
	\caption{ScaMaha tool characteristics.}
	\begin{tabular}{|c|c|c|c|c|c|c|c|c|c|} \hline
		\rowcolor{LightCyan} \multicolumn{10}{|l|}{ScaMaha code parser}   \\\hline\hline	
		\rowcolor{lightyellow} \multicolumn{10}{|c|}{Code entities (or identifiers)}   \\\hline
		&\multicolumn{3}{c|}{Class} & & \multicolumn{5}{c|} {Method}  \\\cline{2-4}\cline{6-10}
		\begin{sideways}Package         \end{sideways} &
		\begin{sideways}Comment         \end{sideways} &
		\begin{sideways}Super-class     \end{sideways} &
		\begin{sideways}Interface       \end{sideways} &
		\begin{sideways}Attribute       \end{sideways} &
		\begin{sideways}Access level    \end{sideways} &
		\begin{sideways}Comment         \end{sideways} &
		\begin{sideways}Parameter list  \end{sideways} &
		\begin{sideways}Local variable  \color{white}\end{sideways} &
		\begin{sideways}Exception       \end{sideways} 
		\\\hline
		$\times$&$\times$&$\times$&$\times$&$\times$&$\times$&$\times$&$\times$&$\times$&$\times$ \\\hline
		\rowcolor{lightyellow} \multicolumn{10}{|c|}{Code relations (or dependencies)}   \\\hline	
		\multicolumn{9}{|l|}{Inheritance}       &   \ding{51}      \\\hline
		\multicolumn{9}{|l|}{Attribute access}  &   \ding{51}      \\\hline	 
		\multicolumn{9}{|l|}{Method invocation} &   \ding{51}      \\\hline\hline
		\rowcolor{LightCyan} \multicolumn{10}{|l|}{ScaMaha code analyzer}   \\\hline\hline
		\rowcolor{lightyellow}\multicolumn{10}{|c|}{Code metrics}   \\\hline
		\multicolumn{9}{|l|}{Lines of code}       &   $\times$      \\\hline
		\multicolumn{9}{|l|}{Number of packages}  &   $\times$      \\\hline	 
		\multicolumn{9}{|l|}{Number of classes}   &   $\times$      \\\hline
		\multicolumn{9}{|l|}{Number of attributes}&   $\times$      \\\hline
		\multicolumn{9}{|l|}{Number of methods}   &   $\times$      \\\hline	 
		\multicolumn{9}{|l|}{Number of comments}  &   $\times$      \\\hline
		\multicolumn{9}{|l|}{Number of local variables}             &   $\times$ \\\hline
		\multicolumn{9}{|l|}{Number of inheritance relations}       &   $\times$ \\\hline	 
		\multicolumn{9}{|l|}{Number of attribute access relations}  &   $\times$ \\\hline
		\multicolumn{9}{|l|}{Number of method invocation relations} &   $\times$      \\\hline\hline
		\rowcolor{LightCyan} \multicolumn{10}{|l|}{ScaMaha code visualizer}   \\\hline\hline
		\rowcolor{lightyellow}\multicolumn{10}{|c|}{Code visualizations}   \\\hline
		\multicolumn{9}{|l|}{Code organization}          &  \ding{51}  \\\hline
		\multicolumn{9}{|l|}{Class inheritance relations}&  \ding{51}  \\\hline
		\multicolumn{9}{|l|}{Method invocation relations}&  \ding{51}  \\\hline
		\multicolumn{9}{|l|}{Polymetric view}            &  \ding{51}   \\\hline
		\multicolumn{9}{|l|}{Tag cloud}                  &  \ding{51}   \\\hline
	\end{tabular}
	\label{tab:ScaMahaChara}
\end{table}

Wettel and Lanza \cite{DBLPWettelL07} have suggested an automatic tool called CodeCity. This tool visualizes software source code as a city metaphor. CodeCity is an interactive, three-dimensional software visualization tool \cite{DBLPMorenoLumbreras23}. CodeCity shows the software code as a city, where the buildings (\resp districts) of the city represent software classes (\resp packages). In CodeCity, building dimensions display values of software metrics, like the number of methods or the number of attributes \cite{DBLPWettelLR11}. While this study presents an automatic tool called ScaMaha, which aims at parsing, analyzing, and visualizing software source code. ScaMaha visualizer generates several graphs regarding several aspects of software code, such as code organization and relations.

The code parser is used in several feature identification (\aka feature location) studies, such as in \cite{ALMsiUVICSR, DBLPAlMsiedeenSHUV13, DBLP:Al-MsiedeenSHUVS13}. It has been used to extract the main source code elements (\eg packages and classes) and relations (\eg attribute access and method invocation) from software products. Also, code parsers are utilized to construct the feature model \cite{ALmsieramimad, Doc.SympAl-Msiedeen1} from OO source code of a collection of software product variants, such as in \cite{AlMsieDeen1445, RaFatbookModel, AlMsiedeenHSUV14, DBLPalllMsiedeenHSUV14, DBLPMeHUV14}. Moreover, code parser are exploited to visualize all software identifier names as tag clouds (\aka word clouds), such as in \cite{AlkkMsiedeenSHUV16, RaFatAhmadDirasat, RaFat2019TagClouds, ALmsiedeen125022}. Furthermore, parsers are employed to identify the traceability links between software source code and its artifacts (\eg software requirements \cite{ALmsiedeen12202111, RaFatRequirement}), such as in \cite{RaFatRT, SalmanSDA12}.

Code parsers are utilized to study OO software evolution based on software identifiers and code relations, such as in \cite{ALmsiedeen154, RaFatImpact}. Furthermore, code parsers are exploited in OO source code summarization studies, such as in \cite{RaFaBlasi}. Moreover, code parsers are utilized in several studies concerning software source code documentation, such as in \cite{FatMsiedeenDocumentation, AlkkMsiedeenSHUV201516, AuLabRafat}. For more information about those parsers, interested readers can refer to the published articles for more details. All these studies show the value of code parsers in the reverse engineering (\resp re-engineering) process.

The Java language has a wide variety of parsers due to its long history of development, popularity, and huge number of applications nowadays. Numerous tools exist that turn software code into a tree-like structure, such as interpreters and compilers. There are several Java parsers for various contexts because there are so many different Java applications, such as Spoon \cite{DBLPPawlakMPNS16}, SrcML \cite{DBLPconfCollardDM13}, and SuperParser \cite{DBLP:Utkin}.

The methods of software source code visualization have become increasingly utilized to support software engineers in software understanding \cite{DBLPFranceseRST16}. In software visualization, some methods aim at displaying software source code in a recognized environment, like a forest \cite{DBLPErraS12} or a city \cite{DBLPWettelLR11}. Another method is to generate what is called a polymetric view \cite{DBLPuLanzaD03}, described as a lightweight visualization technique supplemented with several metrics regarding software code \cite{DBLPLimaMGPGM23}. ScaMaha visualizes different aspects of software source code, such as code organization and relations.

Specialized reverse engineering tools are important and needed these days. Reverse engineering tools are required to deal with the complexity of products and the particular requirements of various tasks, like software comprehension and reuse \cite{DBLPLinsbauer0MAGLE23}. Thus, reverse engineering tools should fit a wide range of circumstances. ScaMaha is a reverse engineering tool for parsing, analyzing, and visualizing software source code. Moose is a well-known reverse engineering tool \cite{DBLPNierstraszDG05}. It began as a research project around 24 years ago. MODMOOSE is the new version of Moose \cite{NicolasAnquetil}. Tool developers can develop specialized reverse engineering tools with MODMOOSE. Moose was based on the Famix meta-model \cite{Ducasse}. MODMOOSE uses FamixNG (a composable meta-model of programming languages), where FamixNG is a redesign of Famix. MODMOOSE utilizes Roassal to script and display interactive graphs \cite{AlexandreBergel}. Roassal is a visualization engine in MODMOOSE. MODMOOSE mainly exploits Roassal to show code entities and their relations in several forms or colors. MODMOOSE uses the MSE file format to describe the source code models \cite{Ducasse}. Where the MSE has been utilized to save FamixNG models. Thus, a software engineer uses an external parser to generate the MSE file of software code, then loads the MSE file into MODMOOSE in order to analyze or visualize it. MODMOOSE is the closest tool to ScaMaha tool.

Lyons \etal \cite{DBLPLyonsBB17} have suggested the lightweight multilingual software analysis method to analyze software systems. They use several code parsers, one for every programming language. The main goal of this work is to create a software engineering tool that addresses large and complex software systems in a lightweight and extensible style. The current version of ScaMaha tool uses only one code parser for the Java language. 

A study of current approaches confirmed the need to offer a comprehensive tool in order to analyze and visualize outdated OO software systems. This work suggests ScaMaha tool, which uses SCA to perform several activities on software code, like code visualization and analysis. This tool accepts only software code and produces a set of code artifacts, which are the code file, code metrics file, and code visualizations. Moreover, this tool helps software engineers understand and maintain legacy software systems. Also, software engineers can easily extend the current version of the tool in order to include other functionalities.

\section{Analyzing and Visualizing OO Source Code via ScaMaha Tool} \label{sec:Approach}

In this study, the author used Java as a target programming language due to its wide adoption in the software engineering field. The Java language is a popular target for both semantic and syntactic code analysis, as well as studies on code visualization \cite{AlkkMsiedeenSHUV201516}, code summarization \cite{RaFaBlasi}, feature location \cite{ALMsiUVICSR, DBLPAlMsiedeenSHUV13}, and re-engineering of software product variants into a Software Product Line (SPL) \cite{AlMsieDeen1445, RaFatbookModel, Mariayyat2014, ALmsiedeen1245}. Application Programming Interfaces (APIs) are available to access and manipulate Java code through Eclipse development tools. Eclipse JDT project provides access to Java code through AST and the Java model. The Java model is displayed as a tree-like structure, and it is used internally to represent each Java project. AST is a comprehensive tree representation of software code.

ScaMaha relies on SCA in order to analyze and visualize software source code. SCA is an important activity to check software code and help discover potential problems early in the cycle of software development before the software system is deployed \cite{DBLP:PrahoLG1, Ilyas947354}. It can discover defects or faults that may be challenging to determine through manual reviews of software code. ScaMaha is a tool for analyzing and visualizing software source code without running it.

\begin{figure}[!htbp]
	\center
	\includegraphics[width=\columnwidth]{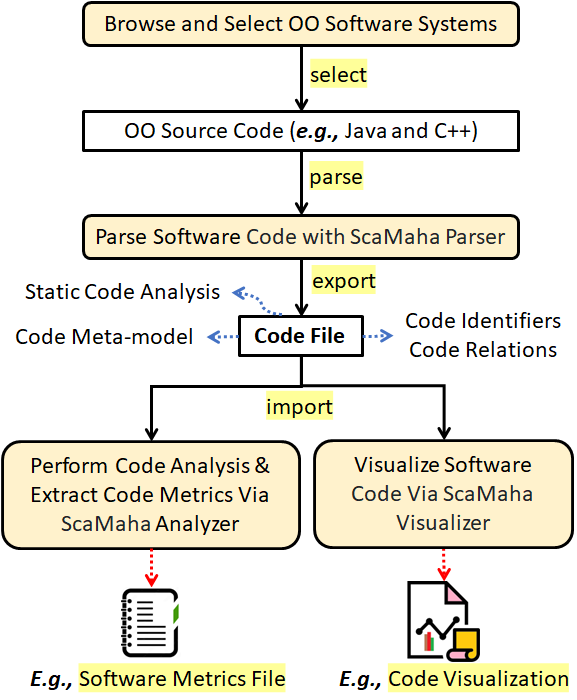}
	\caption{Analyzing and visualizing OO source code with ScaMaha tool.}
	\label{Fig:ScaMaha}
\end{figure}

This study suggests an automatic approach to parse, analyze, and visualize OO software system in a unique manner. The main contribution of this work to the software engineering field is to suggest ScaMaha tool. ScaMaha accepts as input just software source code and produces as output a set of software artifacts, like the code file, software metrics file, and code graphs. ScaMaha tool bases itself on static analysis of software source code. ScaMaha parser extracts all code identifiers and relations. Then, ScaMaha analyzer identifies comprehensive software code metrics, and, at last, ScaMaha visualizer gives a set of graphs for different aspects of software code. Figure \ref{Fig:ScaMaha} shows the process of analyzing and visualizing software source code using ScaMaha tool.

\begin{figure*}[!htbp]
	\center
	\includegraphics[width=\textwidth]{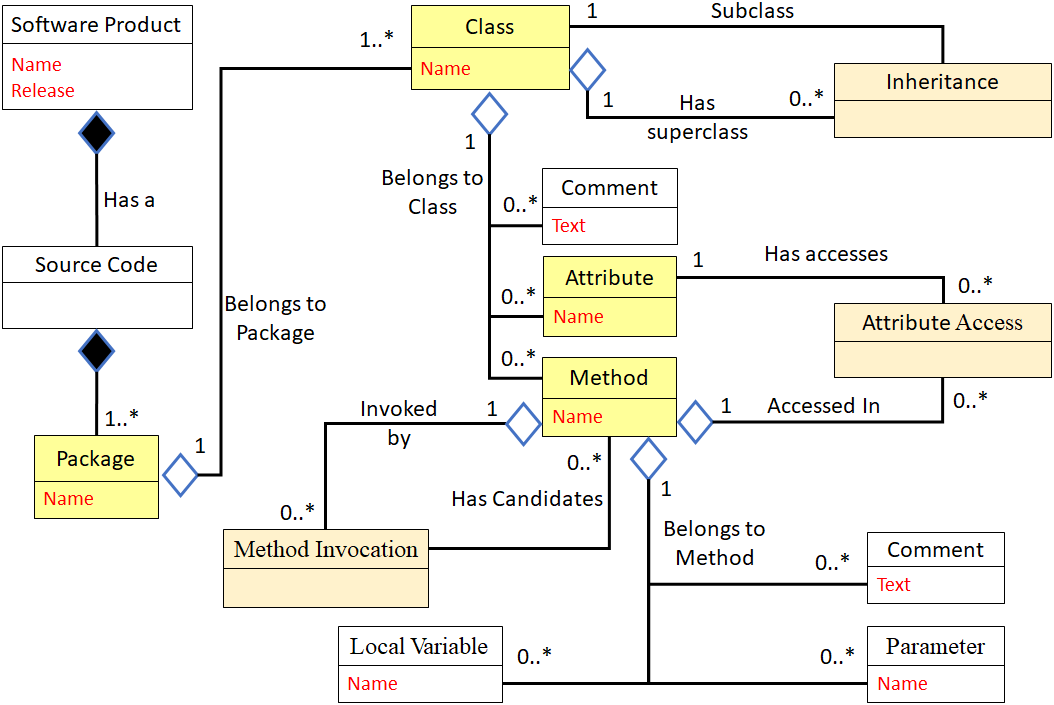}
	\caption{The core of ScaMaha meta-model.}
	\label{Fig:SCAMahaModel}
\end{figure*}

ScaMaha operates on a model of software source code, namely ScaMaha model. To analyze the OO software system, you must first create a model of it using ScaMaha. Figure \ref{Fig:SCAMahaModel} shows the core of ScaMaha meta-model. This meta model shows the main OO software identifiers, like software packages, classes, attributes, and methods. Moreover, this meta model displays the main OO software relationships, such as inheritance, method invocation, and attribute access. The core of ScaMaha involves a language-independent meta-model that can show several OO languages in a uniform style. In most cases, the developer gets sufficient information if he explores the basic types of entities that model an OO software system. These are code identifiers (\eg package) and the relationships (\eg inheritance) between them.

This model shows the majority of OO entities. For instance, it shows that a method has parameters, comments, and local variables. However, while this model doesn't show all OO entities, it is also valuable since, for most practical purposes in the reverse engineering (\resp re-engineering) process, it is all software developers need. Also, ScaMaha model shows that a class (\resp package) has methods (\resp classes), and a method (\resp class) belongs to a class (\resp package). Furthermore, this model shows that an inheritance relation denotes a connection between two classes (\ie the subclass and the superclass). Moreover, ScaMaha model shows that an attribute access relation denotes a connection between method and attribute. Also, it shows that a method invocation relation means a connection between one method and another method.

The first step of code analysis is to explore the software directory (or any workspace) to select OO software that the developers want to analyze. Thus, let's say there is a directory (repository) of OO software projects that developers need to analyze (\eg the source code from the Eclipse workspace). In this case, developers will browse the directory to select a particular software project in order to get a copy of its source code (\cf Figure \ref{Fig:ScaMaha}). If software developers want to analyze Java software, they can use any directory, such as a git repository (\ie GitHub) \cite{DBLPtKangKJ23}.

\begin{figure*}[!htbp]
\begin{lstlisting}[caption={XML format corresponding to ScaMaha meta-model.},language=XML, label=xmlFile]
	<!--By Ra'Fat Al-Msie'deen-->
	<Project Name="----">
		<Packages>
			<Package Name="----">
				<Classes>
					<Class Name="----" AccessLevel="----" isInterface="----" Superclass="----">
						<SuperInterfaces>
							<Interface InterfaceName="----" />
						</SuperInterfaces>
						<Comments>
							<Comment CommentText="----" />
						</Comments>
						<Attributes>
							<Attribute Name="----" AccessLevel="----" Type="----" isStatic="----" />
						</Attributes>
						<Methods>
							<Method Name="----" AccessLevel="----" ReturnType="----" isStatic="----">
								<Parameters NumberOfParameters="----">
									<Parameter Name="----" Type="----" />
								</Parameters>
								<Comments>
									<Comment CommentText="----" />
								</Comments>
								<LocalVariables>
									<LocalVariable Name="----" Type="----" />
								</LocalVariables>
								<AttributeAccesses>
									<Access Name="----" Type="----" HowIsItUsed="----" />
								</AttributeAccesses>
								<MethodInvocations>
									<MethodInvocation Name="----" Arguments="----" />
								</MethodInvocations>
								<MethodAssignments>
									<Assignment LeftHandSide="----" RightHandSide="----" />
								</MethodAssignments>
								<MethodExceptions>
									<Exception ExceptionType="----" />
								</MethodExceptions>
							</Method>
						</Methods>
					</Class>
				</Classes>
			</Package>
		</Packages>
	</Project>
\end{lstlisting}
\end{figure*}

In the second step of ScaMaha, developers parse OO source code to build ScaMaha model (\cf Figure \ref{Fig:SCAMahaModel}). Once developers have a copy of the software source code, they can create a ScaMaha model of software code using ScaMaha parser. To parse OO source code, ScaMaha depends on the static code parser. In this study, the most important entities of source code are considered and parsed, such as packages, classes, methods, and attributes. Also, key relations between main code entities (\ie inheritance, invocation, and access) are considered and parsed. ScaMaha parser generates an XML file of software source code.

In this work, XML is a generic file format that represents ScaMaha code model. Thus, ScaMaha parser converts OO source code to an XML file format. ScaMaha parser produces an XML file for each software product. This file includes all code entities (or identifiers) and relationships (or dependencies) between those entities. Listing \ref{xmlFile} shows the XML format corresponding to ScaMaha meta-model.

As an illustrative example, this study considers the mobile photo software system. ScaMaha used this software to better explain its work. Mobile photo is open-source software that allows users to manipulate photos on their mobile devices \cite{DBLPTizzeiDRGL11}. This study considers the first release of mobile photo software \cite{FigueiredoMobileMedia}. ScaMaha only utilizes the source code of mobile photo as input. Listing \ref{xmlFilMP} shows the parsed code of mobile photo software as an XML file. The parsed XML file includes structural information regarding software code, such as that method “BaseThread” belongs to class “BaseThread” and class “BaseThread” belongs to package “ubc.midp.mobilephoto.core.threads”.

\begin{figure*}[h]
\begin{lstlisting}[caption={The code file of mobile photo software as an XML file (partial) \cite{ScaMaha}.},language=XML, label=xmlFilMP]
	<!--By Ra'Fat Al-Msie'deen-->
	<Project Name="Mobile photo software">
		<Packages>
			<Package Name="ubc">
				<Classes/>
			</Package>
			<Package Name="ubc.midp.mobilephoto.core">
				<Classes/>
			</Package>
			<Package Name="ubc.midp.mobilephoto.core.threads">
				<Classes>
					<Class Name="BaseThread" AccessLevel="public" isInterface="false" Superclass="Object">
						<Comments>
							<Comment CommentText="Start the thread running"/>
						</Comments>
						<Attributes />
						<Methods>
							<Method Name="BaseThread" AccessLevel="public" ReturnType="void" isStatic="false">
								<Parameters NumberOfParameters="0"/>
								<LocalVariables />
								<AttributeAccesses />
								<MethodInvocations>
									<MethodInvocation Name="println" Arguments="[BaseThread:: 0 Param Constructor used ... ]"/>
								</MethodInvocations>
								<MethodAssignments />
								<MethodExceptions />
							</Method>
							<Method Name="run" AccessLevel="public" ReturnType="void" isStatic="false">
								<Parameters NumberOfParameters="0"/>
								<LocalVariables/>
								<AttributeAccesses/>
								<MethodInvocations>
									<MethodInvocation Name="println" Arguments="[Starting BaseThread::run()]"/>
								</MethodInvocations>
								<MethodAssignments/>
								<MethodExceptions/>
							</Method>
						</Methods>
					</Class>
				</Classes>
			</Package>
		</Packages>
	</Project>
\end{lstlisting}
\end{figure*}

ScaMaha exploits Eclipse JDT and AST to parse software systems written in Java. Several studies utilized Eclipse AST to access, read, and manipulate the software code. XML enables easy interchange of metadata between several tools in diverse environments (\cf Figure \ref{Fig:useOfSCAMaha}). Moreover, XML is a human-readable format. The XML structure matches the author's requirements for the representation and description of ScaMaha model.

The chosen way to load a model in ScaMaha is through an XML file. XML is a compact, simple, and robust format. This study exploits XML to represent the core of ScaMaha model. In order to analyze OO source code, software developers need to load the code model as an XML file into ScaMaha tool. ScaMaha analyzer aims at analyzing software code and extracting useful software code metrics. Developers can extend the current work of ScaMaha analyzer by performing other analysis activities on software code. Software developers can use ScaMaha analyzer to obtain several code metrics. For instance, the software LOC and the number of packages. The mined code metrics file gives the software engineer an indication about the size (complexity level) of the software system. In this study, ScaMaha analyzer considers the software metrics presented in Table \ref{tab:ScaMahaMetrics}. Analyzer of ScaMaha can easily extend to include other code metrics.

\begin{table}[!htb]
	\center
	\caption{Software code metrics of ScaMaha code analyzer.}
		\begin{tabular}{|l|c|} \hline \rowcolor{LightCyan}
			Metric                               &  Abbreviation   \\\hline
			Lines of Code                         & LOC             \\\hline			
			Number of Packages                    & NOP             \\\hline			
			Number of Classes                     & NOC             \\\hline			
			Number of Attributes                  & NOA             \\\hline			
			Number of Methods                     & NOM             \\\hline			
			Number of Comments                    & NOCo            \\\hline			
			Number of local variables             & NOLv            \\\hline			
			Number of inheritance relations       & NOIn            \\\hline			
			Number of attribute access relations  & NOAc            \\\hline	
			Number of method invocation relations & NOI             \\		
			\hline
	\end{tabular}
	\label{tab:ScaMahaMetrics}
\end{table}

Source code metrics are measurements utilized to characterize software code. A code metric is a useful quantitative measure extracted from software's source code. Size is the most recognizable metric for software source code. The number of LOC is the easiest method of measuring software size. All code metrics given in Table \ref{tab:ScaMahaMetrics} are static code metrics. Static code metrics are metrics obtained directly from software source code, like the LOC metric. A subgroup of static code metrics are OO code metrics, as they are also metrics obtained from software code itself, such as NOIn, NOI, and NOAc metrics. Listing \ref{MPmetrics} shows the extracted software code metrics from mobile photo software by ScaMaha analyzer.

\begin{lstlisting}[caption={Software code metrics for mobile photo software.},language=XML, label=MPmetrics] 
	<!--By Ra'Fat Al-Msie'deen-->
	<Project ProjectName="Mobile photo software">
		<Metrics>
			<LinesOfCode LOC="1229" />
			<NumberOfPackages NOP="10" />
			<NumberOfClasses NOC="15" />
			<NumberOfAttributes NOA="56" />
			<NumberOfMethods NOM="91" />
			<NumberOfComments NOCo="250" />
			<NumberOfInheritances NOIn="14" />
			<NumberOfInvocations NOI="298" />
			<NumberOfAccesses NOAc="631" />
		</Metrics>
	</Project>
\end{lstlisting}

In this work, the use of visualization speeds up the comprehension of legacy OO source code. By utilizing ScaMaha to analyze and visualize software source code, software developers are able to speed their maintenance, reuse, and comprehension of software products. Source code visualization aims at producing graphical representations (or annotations) of software code in order to help comprehend and analyze it. In the software engineering domain, the code visualization process plays an important role in understanding how large software products work \cite{AlkkMsiedeenSHUV201516}.

\begin{figure}[!htb]
	\center
	\includegraphics[width=\columnwidth]{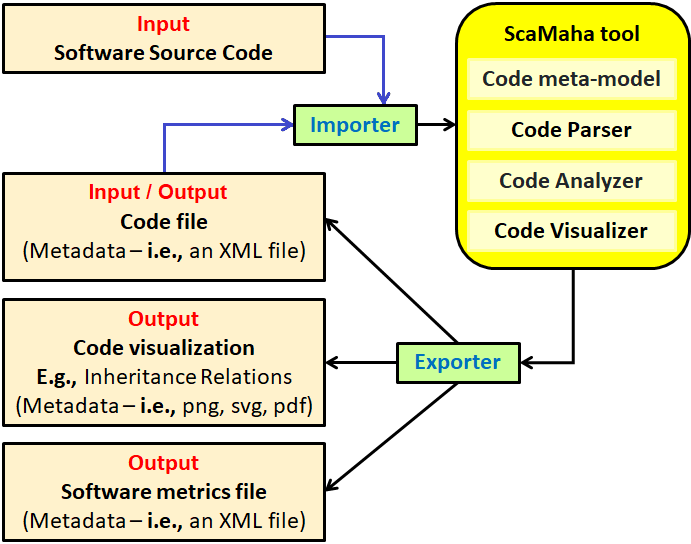}
	\caption{An overview of the core parts of ScaMaha tool.}
	\label{Fig:ALLinOne}
\end{figure}

The main goal of ScaMaha visualizer is to visualize software code entities and relations. All code entities and relations are defined in ScaMaha core meta-model (\cf Figure \ref{Fig:SCAMahaModel}). To visualize software code via ScaMaha visualizer, software developers need to load code files using ScaMaha's importer. Then, the suggested tool will generate different visualizations covering several aspects of software code. ScaMaha exporter stores all code visualizations in the tool workspace. Thus, software developers can explore ScaMaha's workspace to see all the code visualizations. Software developers study code visualizations in order to analyze and understand software products. Figure \ref{Fig:ALLinOne} briefly shows the core parts of ScaMaha tool.

\begin{figure}[!htb]
	\center
	\includegraphics[width=\columnwidth]{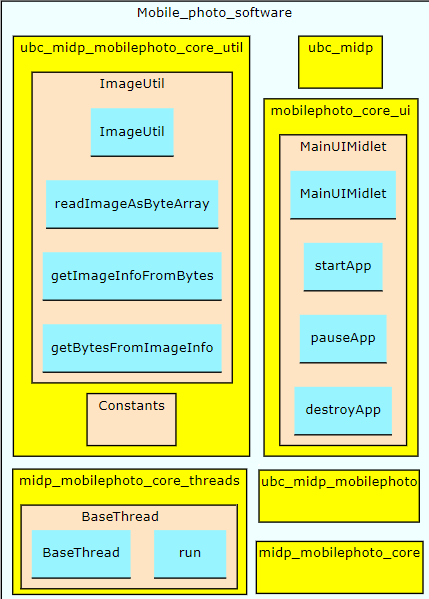}
	\caption{Code organization visualization of mobile photo software (partial) \cite{ScaMaha}.}
	\label{Fig:MPMin}
\end{figure}

ScaMaha visualizes several aspects of software code. The visualization of code organization shows the main code entities as boxes. Code organization visualization represents software packages, classes, and methods. Developers can easily extend the current visualization by adding other entities of code, like software attributes. Figure \ref{Fig:MPMin} shows the code organization visualization of mobile photo software. The main objective of this visualization is to show the organization of code in terms of packages, classes, and methods 

In code organization visualization, the big box represents the whole software code, while other boxes represent main code entities. This visualization shows that the box may contain other boxes. For instance, the package box includes all classes belonging to this package. Also, the class box includes all methods belonging to this class (\cf Figure \ref{Fig:MPMin}).

Furthermore, code organization visualization provides software engineers with structural information about software code. Structural information is the way developers arrange and group software code elements such as package, class, methods, and attributes. In this work, well-structured information about code entities makes software code easier to understand by software engineers (\cf Figure \ref{Fig:MPMin}).

\begin{figure}[!htb]
	\center
	\includegraphics[width=\columnwidth]{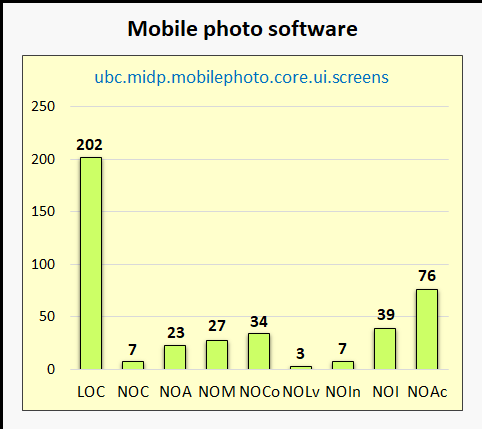}
	\caption{Polymetric view of mobile photo software based on its packages. This view uses the following metrics for each package: LOC, NOC, NOA, NOM, NOCo, NOLv, NOIn, NOI, and NOAc (partial) \cite{ScaMaha}.}
	\label{Fig:PolyMPs}
\end{figure}

ScaMaha also generates a polymetric view of software source code. This view depends on software packages. Where it represents each software package as a box including a set of package metrics. The name of each package is placed on the top of the box. ScaMaha approach considers the following metrics for each package: LOC, NOC, NOA, NOM, NOCo, NOLv, NOIn, NOI, and NOAc. Figure \ref{Fig:PolyMPs} shows a polymetric view of mobile photo software.

\begin{figure}[!htb]
	\center
	\includegraphics[width=\columnwidth]{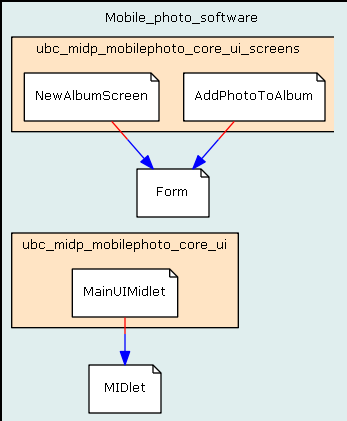}
	\caption{Visualization of class inheritance relations for mobile photo software (partial) \cite{ScaMaha}.}
	\label{Fig:MPooInh}
\end{figure}

In this work, ScaMaha visualizes inheritance relations between software classes. The main goal of inheritance relationships is to minimize code complexity and size. The inheritance relationship gives an indication of the strong connection between software classes. The visualization of class inheritance relations gives important information about legacy (or outdated) code. This kind of visualization helps software developers when they want to analyze, reuse, understand, and maintain existing code. Figure \ref{Fig:MPooInh} shows ScaMaha visualization of class inheritance relations for mobile photo software.

\begin{figure}[!htb]
	\center
	\includegraphics[width=\columnwidth]{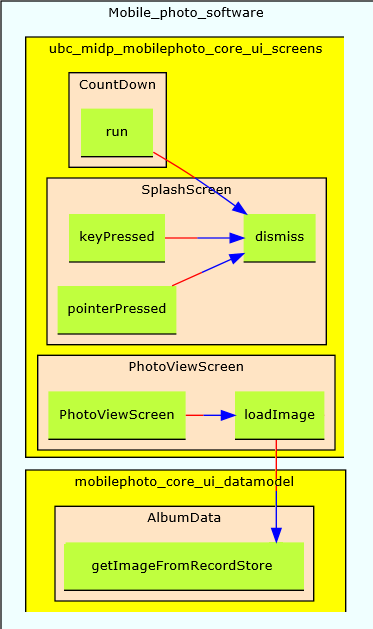}
	\caption{Visualization of method invocation relations for mobile photo software (partial) \cite{ScaMaha}.}
	\label{Fig:MPooInvo}
\end{figure}

In OO language, a method is a function defined inside a class. Thus, a software class may contain several methods. Method communicates with other methods in software via method invocation relations. So, method invocation occurs between software methods when a method calls (or invokes) other methods. Thus, a particular method may invoke other methods of the same class or of different classes (\cf Figure \ref{Fig:MPooInvo}). Usually, a method calls another method by using its name and arguments. Also, methods invoke other methods in order to achieve specific functionality. This study considers method invocation as an important code relationship. ScaMaha visualizes invocation relations across software methods in a perfect way. Software developers explore the extracted visualization in order to comprehend the software code. Figure \ref{Fig:MPooInvo} shows ScaMaha visualization of method invocation relations for mobile photo software.

ScaMaha approach evaluates the produced results using precision, recall, and F-measure metrics \cite{RaFatRT2024}. For source code identifiers and relations, the precision metric is the percentage of correctly recovered identifiers (\resp relations) to the total number of recovered identifiers (\resp relations). While the recall metric is the percentage of correctly recovered identifiers (\resp relations) to the total number of relevant identifiers (\resp relations). Finally, the F-measure metric quantifies a trade-off among precision and recall metrics; thus, it provides a high value just in cases where both recall and precision metrics are high.

All evaluation metrics of ScaMaha have values between zero (0\%) and one (100\%). If the recall metric is equal to 100\%, this means that all relevant identifiers (or relations) are recovered. If the precision metric is equal to 100\%, this means that all retrieved identifiers (or relations) are relevant. If the F-measure metric is equal to 100\%, this means that both recall and precision are high (100\%) \cite{AlMsieDeen1445}. 

In mobile photo software, the proposed approach returns 15 classes from the software code. In this case, the values of all evaluation metrics are equal to 100\% since the software code actually contains only 15 classes. Furthermore, for the method invocation relations, ScaMaha returns 298 relations from the software code. In this circumstance, the values of all evaluation metrics are equal to 100\% since the software code actually contains 298 invocation relations. For software code metrics, the retrieved value of each metric is totally correct. Thus, the values of all evaluation metrics are equal to 100\%. For code visualizations, all types of code visualizations have high accuracy. For instance, in mobile photo software, the visualization of class inheritance relations obviously shows the 14 inheritance relations. In this case, the values of all evaluation metrics are equal to 100\%. Mobile photo is well-documented software \cite{FigueiredoMobileMedia}. Thus, all code entities, metrics, and relations are known in advance. Therefore, the available software documentation helps in comparing ScaMaha results against it.

~\\\textbf{ScaMaha tool in a nutshell:} ScaMaha is a tool for software code analysis and visualization. Also, it is a creative tool to navigate, analyze, and visualize OO software systems. Moreover, ScaMaha helps software developers to cheaply construct custom analyses of software code. ScaMaha implementation is based on Java, and it's an open-source tool. ScaMaha can't be used to analyze the dynamic execution of software. So, it uses SCA, which is a method of exploring the software code without running it. The current version of ScaMaha is shipped with a Java code parser. Also, this tool uses the XML structure to represent software code. Thus, XML is a file format that describes ScaMaha meta-model. XML is generated by an internally provided code parser.
In this work, ScaMaha meta-model is an abstract representation of software source code. In general, it presents software code entities and relations. The code meta-model of ScaMaha is a generic model, and it may describe software systems written in Java and C++ (\ie OO languages). Also, ScaMaha tool includes a code analyzer and visualizer. The code analyzer (\resp visualizer) uses the generated XML file from the code parser. ScaMaha analyzer (\resp visualizer) generates software code metrics (\resp visualizations). ScaMaha visualizer is an integral part of the tool that allows developers to visualize dependencies between classes and methods. In addition, ScaMaha visualizer gives a polymetric view of software based on its packages. Also, it reveals, in a unique visualization, the organization (or structure) of software code. Software engineers view and explore the generated code artifacts via ScaMaha tool to understand and analyze the chosen software systems.

\section{Experimentation}\label{sec:Exper.}

This section presents the case studies used in this work. Also, it presents and discusses the obtained results. Furthermore, it mentions threats to the validity of ScaMaha approach.

ScaMaha runs experiments on several software systems, such as drawing shapes \cite{RaFatRT}, mobile photo \cite{DBLPTizzeiDRGL11}, health watcher \cite{DBLP:PaivaDFS17}, Rhino \cite{Mozirhino}, and ArgoUML \cite{DBLP:MoreiraAMF22}. Drawing shapes software allows users to draw several kinds of shapes, such as ovals and lines \cite{drawingshapes}. Mobile photo is introduced and used in Section \ref{sec:Approach}. The health watcher software is a web-based information system that enables users to register complaints about health issues. In this study, the experiment ran on the last version of health watcher software (\ie version 10) \cite{HWsoftware}. Moreover, Rhino is software for JavaScript developed in Java. In this study, the experiment ran on version 1.7R2 of Rhino. While ArgoUML is an open-source software written in Java \cite{ArgoUML}. It is widely utilized for designing software systems in Unified Modeling Language (UML). It is a large and complex software system (\ie 271690 LOC). In this work, all experiments are executed on a 2.40 GHz Intel Core i7 PC with 8 GB of RAM. Table \ref{tab:Results} briefly presents the results obtained from all experiments.

\begin{table*}[!htb]
	\centering
	\caption{ScaMaha results for all experiments (\ie case studies).}
	\scalebox{0.94}{
		\begin{tabular}{|c|l|c|c|c|c|c|c|c|c|c|} \hline 
			\rowcolor{LightCyan}
			ID&Case study& Software  &\multicolumn{2}{c|}{Software artifacts}& Execution & \multicolumn{5}{c|}{Visualization}   \\\cline{4-5}\cline{7-11} 
			\rowcolor{LightCyan}
			&     &size& Code  & Metrics file & time (in \textit{ms}) &Code$^a$&Class$^b$&Method$^c$&Polymetric$^d$ & Cloud$^e$ \\\hline
			1 & Drawing shapes& Small &\ding{51}& \ding{51}  & 2095   & $\times$ & $\times$ &$\times$&$\times$ &$\times$ \\\hline
			2 & Mobile photo  & Medium&\ding{51}& \ding{51}  & 2850   & $\times$ & $\times$ &$\times$&$\times$ &$\times$ \\\hline
			3 & Health watcher& Medium&\ding{51}& \ding{51}  & 5031   & $\times$ & $\times$ &$\times$&$\times$ &$\times$ \\\hline
			4 & Rhino         & Medium&\ding{51}& \ding{51}  & 7965   & $\times$ & $\times$ &$\times$&$\times$ &$\times$\\\hline 
			5 & ArgoUML       & Large &\ding{51}& \ding{51}  & 31698  & $\times$ & $\times$ &$\times$&$\times$&$\times$ \\\hline 
	\end{tabular}}
	\label{tab:Results}
	\\\footnotesize{$^a$ Code organization, $^b$ Class inheritance relations, $^c$ Method invocation relations, $^d$ Polymetric view, $^e$ Tag cloud.}
\end{table*}

In this work, all important information from the software source code is parsed to form the software code file. Listing \ref{partialDSA} illustrates the parsed code from the drawing shapes software as an XML file.

\begin{lstlisting} [caption={The code file of drawing shapes software as an XML file (partial) \cite{ScaMaha}.},language=XML, label=partialDSA]
	<!--By Ra'Fat Al-Msie'deen-->
	<Project Name="Drawing shapes software">
		<Packages>
			<Package Name="Drawing.Shapes.coreElements">
				<Classes>
					<Class Name="MyLine" AccessLevel="public" isInterface="false" Superclass="MyShape">
						<SuperInterfaces />
						<Comments>
							<Comment CommentText="Class that declares a line object" />
						</Comments>
						<Attributes />
						<Methods>
							<Method Name="draw" AccessLevel="public" ReturnType="void" isStatic="false">
								<Parameters NumberOfParameters="1">
									<Parameter Name="g" Type="Graphics" />
								</Parameters>
								<LocalVariables />
								<AttributeAccesses>
									<AttributeAccess Name="g" Type="Graphics" HowIsItUsed="g.setColor(getColor())" />
								</AttributeAccesses>
								<MethodInvocations>
									<MethodInvocation Name="setColor" Arguments="[getColor()]" />
								</MethodInvocations>
								<MethodAssignments />
								<MethodExceptions />
							</Method>
						</Methods>
					</Class>
				</Classes>
			</Package>
		</Packages>
	</Project>
\end{lstlisting}

In this work, the different sizes of software systems show the scalability of ScaMaha to work with such systems (\ie small, medium, and large systems). Moreover, all software systems are well documented. Thus, the results of ScaMaha are measurable. In addition, all case studies are well known and employed to assess many approaches in the field of this study (\ie SCA).

During software maintenance, software engineers consume a considerable amount of time analyzing legacy software system source code in order to understand it. Furthermore, the cost of software maintenance accounts for 50\% to 75\% of the total cost of the software product \cite{DBLPr0084424}. Thus, comprehending software source code is one of the most challenging activities in the maintenance (\resp comprehension) of software products. Listing \ref{DSmetrics} shows the extracted code metrics from drawing shapes software.

\begin{lstlisting}[caption={Software code metrics for drawing shapes software.},language=XML, label=DSmetrics]
	<!--By Ra'Fat Al-Msie'deen-->
	<Project ProjectName="Drawing shapes software">
		<Metrics>
			<LinesOfCode LOC="213" />
			<NumberOfPackages NOP="4" />
			<NumberOfClasses NOC="6" />
			<NumberOfAttributes NOA="16" />
			<NumberOfMethods NOM="29" />
			<NumberOfComments NOCo="112" />
			<NumberOfInheritances NOIn="6" />
			<NumberOfInvocations NOI="99" />
			<NumberOfAccesses NOAc="125" />
		</Metrics>
	</Project>
\end{lstlisting}

Listing \ref{ArgoUMLmetrics} shows the obtained code metrics from ArgoUML software. Software developers can easily detect that ArgoUML is a large software system using the extracted code metrics (\eg the number of software classes is equal to 1939).

\begin{lstlisting}[caption={Software code metrics for ArgoUML software.},language=XML, label=ArgoUMLmetrics]
	<!--By Ra'Fat Al-Msie'deen-->
	<Project ProjectName="ArgoUML software">
		<Metrics>
			<LinesOfCode LOC="271690" />
			<NumberOfPackages NOP="90" />
			<NumberOfClasses NOC="1939" />
			<NumberOfAttributes NOA="3977" />
			<NumberOfMethods NOM="14904" />
			<NumberOfComments NOCo="64929" />
			<NumberOfLocalVariables NOLv="10874" />
			<NumberOfInheritances NOIn="1783" />
			<NumberOfInvocations NOI="56758" />
			<NumberOfAccesses NOAc="76121" />
		</Metrics>
	</Project>
\end{lstlisting}

Usually, software engineers hope to get all code information (\eg software identifiers and relations) to exploit this information in many software engineering activities (\eg maintenance, re-engineering, visualization, documentation, and reverse engineering). Listing \ref{HWMetrics} shows the obtained code metrics from health watcher software.

\begin{lstlisting} [caption={Software code metrics for health watcher software.},language=XML, label=HWMetrics]
	<!--By Ra'Fat Al-Msie'deen-->
	<Project ProjectName="Health watcher software">
		<Metrics>
			<LinesOfCode LOC="8217" />
			<NumberOfPackages NOP="29" />
			<NumberOfClasses NOC="135" />
			<NumberOfAttributes NOA="256" />
			<NumberOfMethods NOM="894" />
			<NumberOfComments NOCo="300" />
			<NumberOfLocalVariables NOLv="602" />
			<NumberOfInheritances NOIn="118" />
			<NumberOfInvocations NOI="2703" />
			<NumberOfAccesses NOAc="4465" />
		</Metrics>
	</Project>
\end{lstlisting}

The obtained metrics for software code give an indication of software size (or complexity level). A software metrics file gives software engineers rapid information about software code, like the number of software methods. Listing \ref{RhinoMetrics} shows the obtained code metrics from Rhino software.

\begin{lstlisting} [caption={Software code metrics for Rhino software.},language=XML, label=RhinoMetrics]
	<!--By Ra'Fat Al-Msie'deen-->
	<Project ProjectName="Rhino software">
		<Metrics>
			<LinesOfCode LOC="139408" />
			<NumberOfPackages NOP="11" />
			<NumberOfClasses NOC="167" />
			<NumberOfAttributes NOA="1854" />
			<NumberOfMethods NOM="2301" />
			<NumberOfComments NOCo="4247" />
			<NumberOfLocalVariables NOLv="4447" />
			<NumberOfInheritances NOIn="146" />
			<NumberOfInvocations NOI="10763" />
			<NumberOfAccesses NOAc="42227" />
		</Metrics>
	</Project>
\end{lstlisting}

\begin{figure*}[!htb]
	\center
	\includegraphics[width=\textwidth]{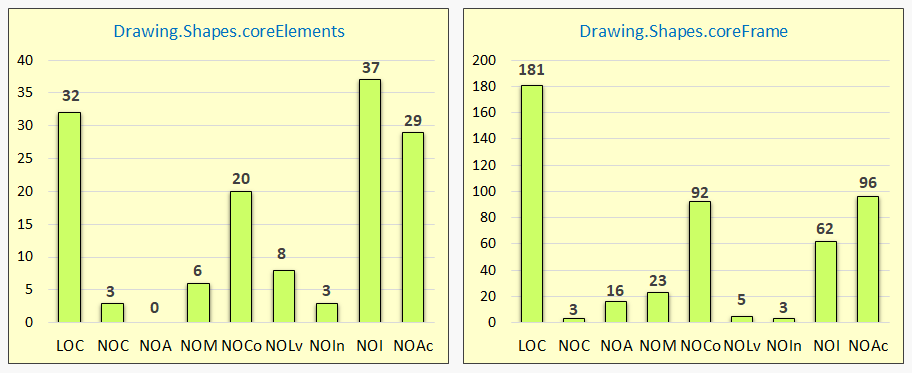}
	\caption{Polymetric view of drawing shapes software based on its packages. This view uses the following metrics for each package: LOC, NOC, NOA, NOM, NOCo, NOLv, NOIn, NOI, and NOAc.}
	\label{Fig:polymetric}
\end{figure*}

One of the main contributions of ScaMaha approach is to give a polymetric view of software packages. In this study, polymetric view is a lightweight visualization method of software source code \cite{DBLPuLanzaD03}. Polymetric views supplemented with unique metrics about software source code \cite{DBLPFranceseRST16}. Polymetric views assist software engineers in understanding the complexity level of software code in the reverse engineering process. Figure \ref{Fig:polymetric} shows a polymetric view of drawing shapes software based on its packages. This view uses the following metrics for each package: LOC, NOC, NOA, NOM, NOCo, NOLv, NOIn, NOI, and NOAc. Figure \ref{Fig:polymetric} shows that drawing shapes software consists of two packages. Also, several metrics regarding each package are given in Figure \ref{Fig:polymetric}.

\begin{figure}[H]
	\includegraphics[width=\columnwidth]{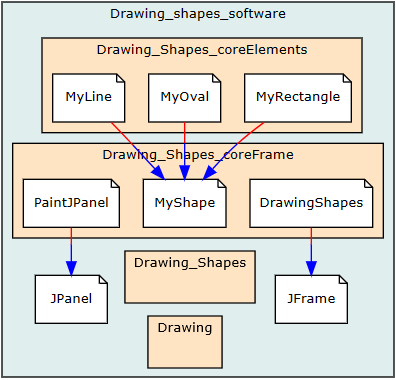}
	\caption{Visualization of class inheritance relations for drawing shapes software.}
	\label{Fig:DSInhNr}
\end{figure}

Software visualization is an important activity in the software engineering domain. Source code visualization is a real implementation of the quote, “A picture is worth a thousand words.”. Code visualization gives developers better information compared to textual code information. ScaMaha visualizations are a real reflection of software code. ScaMaha visualizes code identifiers and relations correctly. The visualizer accepts as input the code file and generates as output a collection of code visualizations. Figure \ref{Fig:DSInhNr} shows ScaMaha visualization of class inheritance relations from drawing shapes software.

Figure \ref{Fig:DSInhNr} gives different views of software code, where it presents structural information (\eg MyLine class belongs to the coreElements package), identifier names (\eg MyShape class), and inheritance relations (\eg MyLine class inherits attributes and methods of MyShape class) from drawing shapes software.

\begin{figure}[!htb]
	\includegraphics[width=\columnwidth]{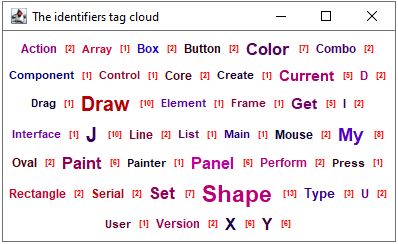}
	\caption{Tag cloud extracted from the drawing shapes software using ScaMaha.}
	\label{Fig:allIdentifiers}
\end{figure}

Figure \ref{Fig:allIdentifiers} represents a tag cloud generated from the source code of the drawing shapes software by ScaMaha. It displays key software identifiers such as package names, class names, attribute names, and method names, with larger tags indicating higher frequency and importance. Tag frequencies are shown in red within square brackets.

The obtained results from all experiments show that all evaluation metrics (\ie precision, recall, and F-measure) appear high (\ie 100\%) for all artifacts extracted from software source code, such as code and metrics files and code visualizations. This means that all resulted artifacts (\eg code visualizations) from software code are relevant and correct. In this study, the author uses two resources to evaluate the obtained results. The first is the available software documents, and the second is the manual review of software code.

Results show the ability of ScaMaha to retrieve all software identifiers (\eg software classes and attributes) from any software system. Also, the results show that ScaMaha is able to retrieve all code comments (\ie class and method comments). Moreover, ScaMaha tool is capable of retrieving all elements of the method body (\ie parameter list, local variable, method invocation, and attribute access). Thus, ScaMaha guarantees that software developers will not lose any code identifier or relation from the software code.

\begin{figure*}[!htb]
	\centering
	\includegraphics[width=\textwidth]{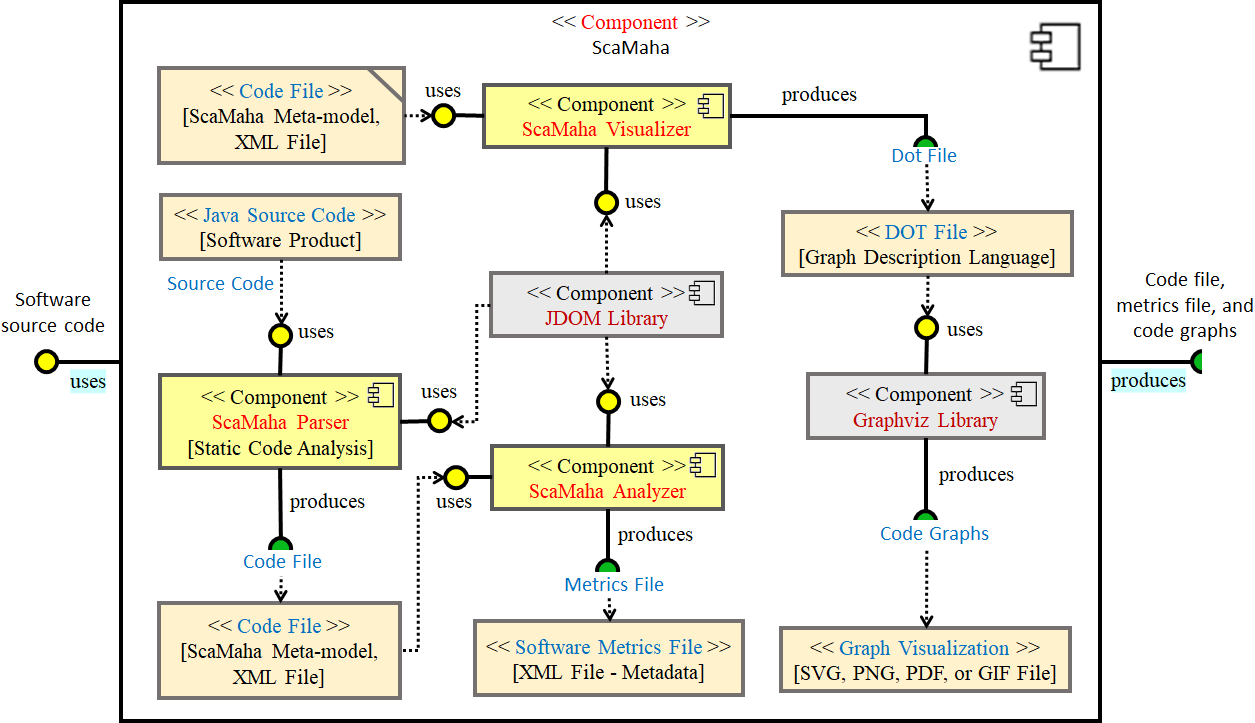}
	\caption{An architectural view of ScaMaha tool in a simplified structure.}
	\label{Fig:architecture}
\end{figure*}

ScaMaha tool consists of three basic components, which are the code parser, analyzer, and visualizer. Moreover, ScaMaha exploits and reuses two components, which are the JDOM \cite{JDOM} and Graphviz libraries \cite{Graphviz}. ScaMaha component accepts software code as input and produces as output a collection of code artifacts (\eg software metrics file). The parser component accepts software code and generates the code file, while the analyzer component accepts the code file and generates a software metrics file. Finally, the visualizer component receives the code file as input and creates several code visualizations as outputs. Figure \ref{Fig:architecture} shows an architectural view of ScaMaha tool in a simplified structure.

By viewing the extracted software metrics file, the software developers can easily determine the size and complexity level of the analyzed software. For instance, based on the mined code metrics file, drawing shapes software is considered as a small software (\cf Listing \ref{DSmetrics}), while ArgoUML is considered as a large and complex software system (\cf Listing \ref{ArgoUMLmetrics}). The software metrics file includes unique code metrics, like the number of inheritance relations between software classes. The originality of ScaMaha analyzer is that it exploits all important code metrics to determine the size and complexity of software code.

Table \ref{tab:Results} shows the results obtained for each case study using ScaMaha tool. It shows the extracted code file (\resp code metrics file) for each case study. Also, it shows the mined code visualizations for each case study. Moreover, the time needed to parse, analyze, and visualize each case study in \textit{ms} is given in Table \ref{tab:Results} (\ie execution time). Results show the ability of ScaMaha to work with different software system sizes (\ie small and large software systems). Also, results show the ability of ScaMaha visualizer to provide several visualizations of software code. Thanks to ScaMaha tool, software developers can easily parse, analyze, and visualize OO software systems. Complete results of ScaMaha experiments are available on the tool's webpage \cite{ScaMaha}.

\begin{figure}[!htbp]
	\includegraphics[width=\columnwidth]{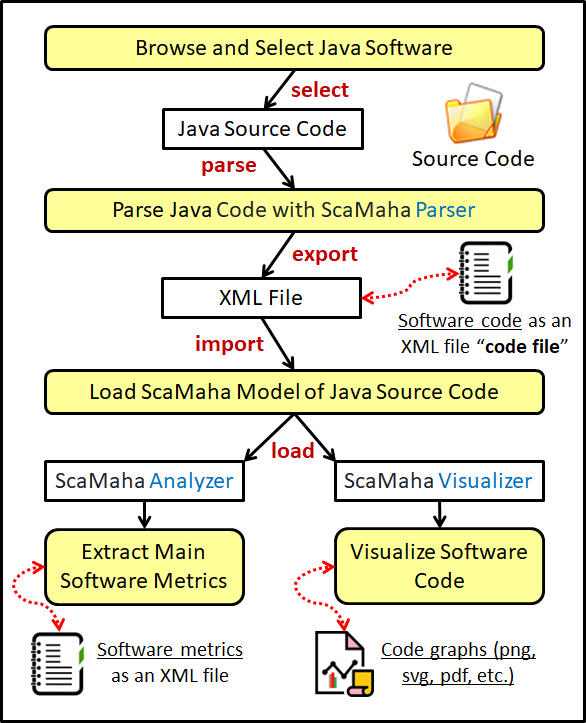}
	\caption{The Java implementation of ScaMaha tool.}
	\label{Fig:futureWork}
\end{figure}

In conclusion, software engineers can use ScaMaha tool for several tasks during their daily work, like parsing, analyzing, and visualizing software source code. ScaMaha tool targets several areas of the software engineering field, like software comprehension, visualization, and maintenance \cite{DBLPButtAM22}. The current version of the tool only works with software systems written entirely in Java (\cf Figure \ref{Fig:futureWork}). In this case, there is a threat to the validity of the prototype, which restricts the ability of ScaMaha implementation to work only with software systems written in Java. To solve this problem, there is a need to develop a parser for each programming language, like C++. But the parser should generate a code file identical to the XML file that is used by ScaMaha's tool. Then, developers can import a code file, load it into ScaMaha tool, and continue to perform other tasks (\ie code analysis and visualization).

\section{Conclusion and Future Work}\label{sec:conclusion}

This paper has presented ScaMaha, a tool for parsing, analyzing, and visualizing OO source code like Java. ScaMaha is designed to prevent software engineers from wasting their resources, like effort and time, on manual review of software source code in order to understand it. The main goal of ScaMaha tool is to assist software engineers in the process of understanding complex and large-sized software systems. Various categories of users, such as scholars in the field of software analysis and tool creators, will exploit ScaMaha tool in their works.

In summary, ScaMaha extracts code identifiers and relationships, generates a metrics file with statistical data, and produces unique graphs to effectively visualize software code. ScaMaha had been validated and evaluated on several case studies, including drawing shapes, mobile photo, health watcher, rhino, and ArgoUML software. The results of the experiments show the capacity of ScaMaha to recover all software artifacts in an efficient and accurate manner. The evaluation metrics of ScaMaha, like precision and recall, show the accuracy of ScaMaha in parsing, analyzing, and visualizing software source code, as all source code artifacts were correctly obtained.

Regarding ScaMaha's future work, the author plans to extend the current tool by developing a comprehensive tool's parser for all OO languages, like Java and C++. Moreover, additional experimental tests can be performed to verify ScaMaha contributions utilizing open-source and industrial software systems. Also, the author plans to conduct a comprehensive survey regarding current approaches that relate to ScaMaha contributions. Finally, the author of ScaMaha tool plans to extend the current work by developing a general tool for performing various kinds of analyses and visualizations of any OO software system.
\bibliographystyle{IEEEtran}
\bibliography{references}
\end{document}